\documentclass[sigconf]{acmart}
\AtBeginDocument{%
  }

\copyrightyear{2025}
\acmYear{2025}
\setcopyright{acmlicensed}\acmConference[CIKM '25]{Proceedings of the 34th ACM International Conference on Information and Knowledge Management}{November 10--14, 2025}{Seoul, Republic of Korea}
\acmBooktitle{Proceedings of the 34th ACM International Conference on Information and Knowledge Management (CIKM '25), November 10--14, 2025, Seoul, Republic of Korea}
\acmDOI{xx/xx}
\acmISBN{979-8-4007-2040-6/2025/11}

\usepackage{multirow} 
\usepackage{pdflscape}
\usepackage[ruled,vlined]{algorithm2e}
\usepackage{enumitem}

\begin{document}

\title{Stratified Expert Cloning for Retention-Aware Recommendation at Scale}

\author{Chengzhi Lin}
\affiliation{%
  \institution{Kuaishou Technology}
  \city{Beijing}
  \country{China}
}\email{1132559107@qq.com}

\author{Annan Xie}
\affiliation{%
  \institution{Peking University}
  \city{Beijing}
  \country{China}
}
\email{2301210470@stu.pku.edu.cn}

\author{Shuchang Liu, Wuhong Wang, Chuyuan Wang, Yongqi Liu, Han Li}
\affiliation{%
  \institution{Kuaishou Technology}
  \city{Beijing}
  \country{China}
}
\email{{liushuchang, wangwuhong}@kuaishou.com}
\email{{wangchuyuan, liuyongqi}@kuaishou.com}
\email{lihan@kuaishou.com}

\renewcommand{\shortauthors}{Chengzhi Lin et al.}


\begin{abstract}
 User retention is critical in large-scale recommender systems, significantly influencing online platforms' long-term success. Existing methods typically focus on short-term engagement, neglecting the evolving dynamics of user behaviors over time. Reinforcement learning (RL) methods, though promising for optimizing long-term rewards, face challenges like delayed credit assignment and sample inefficiency.

We introduce Stratified Expert Cloning (SEC), an imitation learning framework that leverages abundant interaction data from high-retention users to learn robust policies. SEC incorporates: 1) multi-level expert stratification to model diverse retention behaviors; 2) adaptive expert selection to dynamically match users with appropriate policies based on their state and retention history; and 3) action entropy regularization to enhance recommendation diversity and policy generalization.

Extensive offline evaluations and online A/B tests on major video platforms (Kuaishou and Kuaishou Lite) with hundreds of millions of users validate SEC’s effectiveness. Results show substantial improvements, achieving cumulative lifts of 0.098\% and 0.122\% in active days on the two platforms respectively, each translating into over 200,000 additional daily active users.
\end{abstract}

\ccsdesc[500]{Information systems~Recommender systems}

\keywords{Retention, Stratified Expert Cloning, Imitation Learning}


\maketitle
\section{Introduction}
\begin{figure}[t]
    \centering
    \includegraphics[width=1\linewidth]{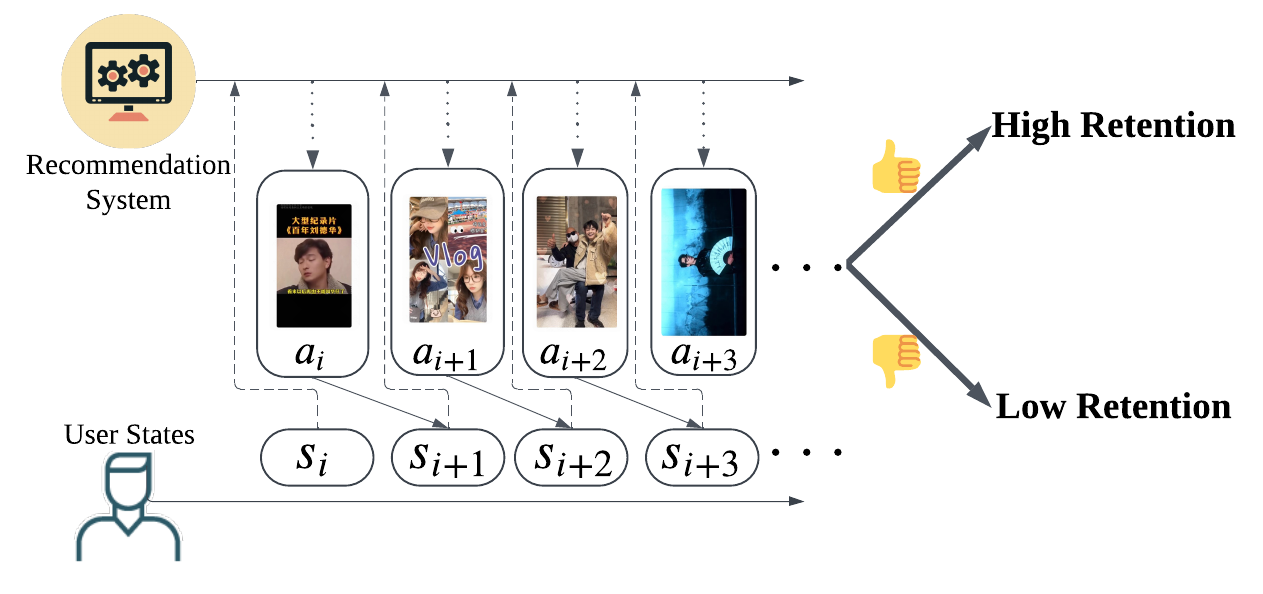}
    \caption{Schematic illustration of our recommendation system based on user state sequences. The system tracks user state transitions $(s_{i}, s_{i+1}, ...)$ and makes content recommendations, i.e., action sequences $(a_{i}, a_{i+1}, ...)$, with the ultimate goal of high user retention reward.}
    \label{fig:scene}
    \vspace{-0.3cm}
\end{figure}
In the era of information abundance, recommender systems are central to online platforms, significantly influencing user experience and content consumption~\cite{koren2009matrix,shi2017long,1511.06939,quadrana2018sequence}. Among the many challenges faced by recommender systems, user retention—defined as sustaining long-term user engagement—is critical for sustainable growth. However, it remains difficult due to the constantly evolving nature of user preferences and behaviors, as illustrated in Figure~\ref{fig:scene}.

Traditional methods such as collaborative filtering and content-based approaches~\cite{cf_1, cf_2, cf_3, cf_4} focus on short-term engagement metrics like click-through rate, often overlooking the dynamics of long-term user retention. While recent advances in reinforcement learning (RL)~\cite{park2013accuracy, PLUR, 2404.03637, gfn} offer promising solutions for sequential decision-making in recommender systems, they still face significant challenges—including delayed rewards, large action spaces, and evolving user behaviors—that hinder effective credit assignment, sample efficiency, and reward design.

To address these limitations, we turn to imitation learning, leveraging expert demonstration data—particularly from high-retention users—to bypass the inefficiencies of RL exploration. We propose Stratified Expert Cloning (SEC),  an imitation learning framework that stratifies expert users by retention levels and learns specialized behavior policies tailored to each group.

SEC encompasses three core technical innovations designed to optimize long-term user retention: (1) multi-level expert stratification, which segments expert users into distinct retention-based hierarchies and learns specialized recommendation policies for each level, enabling nuanced behavioral modeling across the retention spectrum; (2) adaptive expert selection, which dynamically assigns users to the most appropriate expert policy based on their current state embedding and historical interaction patterns, providing personalized recommendation strategies; and (3) action entropy regularization, which promotes recommendation diversity through nuclear norm maximization in the action space, preventing policy collapse while enhancing exploration and generalization across user interests. Together, these components form a comprehensive framework that effectively leverages expert behaviors to improve user retention in large-scale recommender systems.

Through comprehensive offline and extensive online A/B tests on two major video platforms, SEC substantially improves user retention compared to state-of-the-art methods, highlighting its effectiveness and practicality in large-scale recommendation scenarios.

The contributions of our work are summarized as follows:
\begin{itemize}[leftmargin=*]
 \item We propose SEC, a novel imitation learning framework for hierarchical behavior cloning from expert users to optimize long-term retention.
 \item We introduce adaptive expert selection based on user state and history to personalize recommendations.
 \item We employ action entropy regularization to enhance recommendation diversity and generalization.
 \item We validate SEC's superiority through extensive experiments and online evaluations on large-scale platforms.
\end{itemize}

\begin{figure*}
    \centering
    \includegraphics[width=0.95\linewidth]{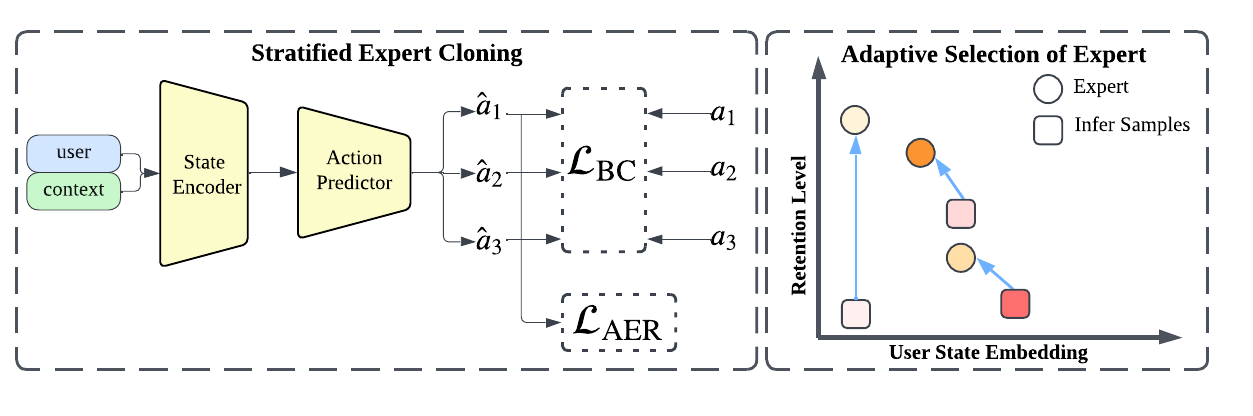}
    \vspace{-0.7cm}
    \caption{Overview of the Stratified Expert Cloning (SEC) framework. Left: During training, the framework learns specialized recommendation policies through behavior cloning from users at multiple high-retention expert levels. The state encoder processes user context and state information, while multiple expert policies are trained simultaneously to capture recommendation strategies from different levels of expert users. Right: At inference time, the framework adaptively selects the most appropriate expert policy based on the current user's state embedding and historical retention level. 
}
\label{fig:method}
\vspace{-0.2cm}
\end{figure*}

\section{Related Work}
\subsection{User Retention in Recommendation Systems}

User retention is crucial for recommender systems as long-term engagement generates higher value than short-term interactions. Traditional methods such as collaborative filtering and matrix factorization primarily focus on immediate engagement metrics (e.g., click-through rates) but often struggle with modeling temporal dynamics necessary for long-term retention \cite{koren2009matrix,shi2017long}. Recent advances like DT4Rec \cite{zhao2023user} utilize Decision Transformers to optimize retention by modeling interaction sequences with discretized reward prompts. Similarly, IURO \cite{ding2023interpretable} explicitly models interpretable factors influencing retention, while Generative Flow Networks \cite{gfn} treat user retention as probabilistic flows over sessions, effectively handling sparse and delayed signals.

Moreover, recommendation diversity methods like Maximal Marginal Relevance (MMR) \cite{carbonell1998use,park2013accuracy} demonstrate that diversified recommendations help maintain user interest, reducing recommendation fatigue and improving retention. Empirical studies \cite{kwon2020art} confirm diversity significantly boosts long-term engagement and purchase rates.

\subsection{Imitation Learning}

Reinforcement Learning (RL), though effective for long-term optimization, faces challenges like high sample complexity and risks from exploration, particularly in recommender systems. Imitation Learning (IL) \cite{ng2000algorithms, im_survey} addresses these issues by learning from expert behaviors, eliminating costly trial-and-error interactions.

Behavioral Cloning (BC), an IL method notable for its simplicity, learns directly from expert demonstrations as supervised learning \cite{bc_c1,bc_c2}. BC has proven effective in diverse domains, including large language model~\cite{chatgpt,dp0,kto}, robotics~\cite{ibc,r2} and autonomous driving~\cite{av, av2}.

Advanced IL methods like Inverse Reinforcement Learning (IRL) \cite{ng2000algorithms} and Generative Adversarial Imitation Learning (GAIL) \cite{ho2016generative} infer reward functions and align agent-expert distributions respectively, enhancing IL's generalizability. Models such as DQfD \cite{hester2018deep} and ChauffeurNet \cite{1812.03079} exemplify IL's versatility across complex tasks.

In recommendation contexts, IL (particularly BC) is promising for optimizing user retention by replicating behaviors of high-retention users. Our proposed Stratified Expert Cloning (SEC) framework enhances BC by learning specialized policies from stratified expert users, effectively capturing diverse behaviors and promoting sustained user engagement.

\section{PRELIMINARIES}

\subsection{Problem Formulation}

We consider a large-scale recommender system where a set of users $\mathcal{U}$ interacts with a set of items $\mathcal{I}$ over multiple sessions. The goal is to provide personalized recommendations that maximize long-term user retention, defined as the likelihood of a user remaining active on the platform over an extended period. User retention is often measured by metrics such as the number of active days within a given time window or the time until the next return to the platform.

Let $\tau = \{(s_1, a_1), (s_2, a_2), \ldots, (s_T, a_T)\}$ denote a user trajectory, where $s_t$ represents the user state at time step $t$ encoding relevant user information, such as demographics, preferences, and recent interactions, while the action $a_t$ represents the recommended item or set of items.

Our objective is to learn a recommendation policy $\pi: \mathcal{S} \rightarrow \mathcal{A}$ that maximizes the expected long-term user retention:

\begin{equation}
\max_{\pi} \mathbb{E}_{u \sim \mathcal{U}} [R_u | \pi].
\end{equation}

\subsection{Challenges in Optimizing Long-term User Retention}

Optimizing long-term user retention in real-world recommender systems presents several technical challenges:
\begin{itemize}[leftmargin=*]
\item \textbf{Delayed Impact and Credit Assignment}: The impact of recommendations on user retention is often delayed, making it difficult to attribute the retention outcome to specific actions. For instance, a user may receive a recommendation for a new category, which sparks their interest and leads to increased engagement in the short term. However, the long-term impact of this recommendation on the user's likelihood to remain active on the platform may not be immediately observable. This temporal gap between actions and rewards complicates the credit assignment problem.

\item \textbf{Sample Inefficiency and Exploration}: The vast action space of all possible recommendations, coupled with the need for extensive exploration, can lead to sample inefficiency and suboptimal user experiences. Exploring irrelevant or poor-quality recommendations may frustrate users and hinder retention. Therefore, it is crucial to balance exploration and exploitation in a sample-efficient manner.

\item \textbf{Complex User Preferences and Behaviors}: User preferences and behaviors are dynamic and evolve over time. A user's interests may shift from one  category to another over time, or their sensitivity to certain types of recommendations may change based on their current life stage or external factors. Capturing these complex dynamics and adapting to user-specific patterns is crucial for effective retention optimization.
\end{itemize}

Traditional methods, such as collaborative filtering and content-based approaches~\cite{cf_1, cf_2, cf_3, cf_4}, primarily focus on short-term engagement metrics and fail to capture the long-term effects of recommendations on user retention. Reinforcement learning techniques~\cite{park2013accuracy, PLUR, 2404.03637, gfn}, while promising, face difficulties in credit assignment, sample efficiency, reward definition and exploration when applied to the user retention problem.

\subsection{Imitation Learning for User Retention}
Imitation learning~\cite{im_survey}, specifically behavior cloning, offers a promising approach to address the challenges in optimizing long-term user retention. Behavior cloning is an imitation learning technique that learns a policy to mimic expert actions given states.

\textbf{Motivation and Advantages}: In large-scale recommender systems, we have access to an abundance of interaction data from high-retention users, who can be considered as experts. By leveraging these expert demonstrations, we can learn recommendation policies that directly optimize for long-term user engagement, churn reduction, and user retention. Imitation learning circumvents the difficulties of explicit reward design and costly exploration faced by RL approaches ~\cite{im_survey}. It enables learning directly from successful real-world examples, avoiding the need for trial-and-error in live systems.

\textbf{Behavior Cloning Approach}: Behavior cloning has demonstrated effectiveness in large language model~\cite{chatgpt,dp0,kto}, robotics~\cite{ibc,r2} and autonomous driving~\cite{av, av2}. We aim to adapt it to the user retention problem in recommender systems. Our proposed Stratified Expert Cloning (SEC) framework introduces novel components such as multi-level expert stratification, adaptive expert selection, and action entropy regularization. These innovations enhance vanilla behavior cloning to capture the nuances of user retention dynamics and enable personalized, robust policy learning.

Based on these insights, we propose a novel imitation learning framework specifically designed to address the challenges of optimizing user retention in recommender systems. The following section details our approach.

\section{THE PROPOSED FRAMEWORK}
\subsection{Overview of the SEC Framework}
The Stratified Expert Cloning (SEC) framework optimizes long-term user retention in recommender systems by learning from high-retention expert users. SEC consists of three main components. First, multi-level expert behavior cloning stratifies expert users into $K$ levels based on retention scores and learns a specialized policy $\pi_k$ for each level. Second, action entropy regularization encourages recommendation diversity by incorporating an entropy regularization term in the training objective. Third, an adaptive selection mechanism assigns users to the most appropriate expert level based on their current state and historical interactions. 

Figure \ref{fig:method} illustrates the SEC framework, integrating these components to learn personalized recommendation policies that optimize user retention. The following sections delve into the details of each component, describing the training and inference stages of the SEC framework.

\begin{figure}[t]
    \centering
    \includegraphics[width=0.48\textwidth]{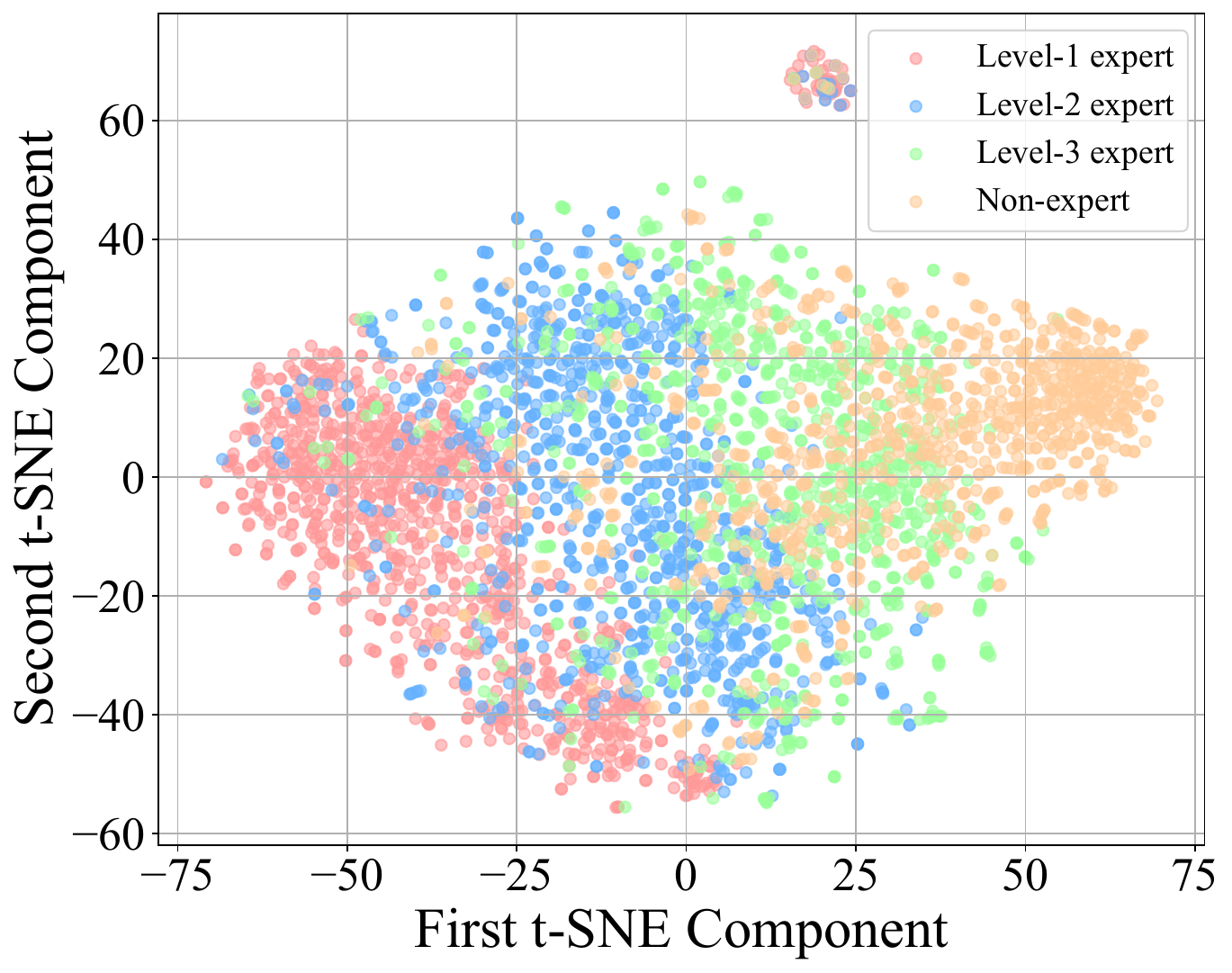}
    \vspace{-0.8cm}
    \caption{T-SNE visualization of user state embeddings from the KuaiRand dataset, categorized by expert levels based on retention patterns. The visualization shows distinct clusters for level-1, level-2, and level-3 experts, with non-expert users distributed across the space. Notably, many non-expert users align more closely with lower-level experts (level-2 and level-3) rather than with the highest-retention experts (level-1), demonstrating that user behaviors exist on a continuum of retention patterns. This empirical finding strongly supports our multi-level expert stratification approach over traditional binary expert/non-expert classifications.}
    \label{fig:user_embeddings}
    \vspace{-0.4cm}    
\end{figure}

\subsection{Training Stage}
The training stage of the SEC framework involves multi-level expert stratification, behavior cloning for each expert level, action entropy regularization, and the optimization of training objectives.

\subsubsection{Multi-Level Expert Stratification}
The SEC framework introduces a multi-level expert stratification approach to capture the nuances in expert user behaviors at different retention levels. Our key insight is that even among expert users with high overall retention, there exist distinct subgroups exhibiting different behavioral patterns. As illustrated in Figure \ref{fig:user_embeddings}, our analysis reveals that non-expert users distribute heterogeneously in the representation space, with many aligning more closely with level-2 or level-3 experts rather than with the highest-retention experts (level-1). This embedding pattern demonstrates that user retention behaviors exist on a continuum rather than in binary states, providing strong empirical evidence for our hierarchical expert stratification approach. By learning specialized policies from multiple expert levels, we can more effectively guide diverse users toward improved retention through personalized recommendation strategies.

 First, we define expert users as those who exhibit high engagement and long-term retention with the recommender system. The criterion for selecting expert users can be based on metrics such as the number of active days, sessions, or interactions within a given time period. Let $\mathcal{U}_E \subseteq \mathcal{U}$ denote the set of expert users.

Then we stratify the expert users into $K$ levels, $\{\mathcal{U}_{E}^1, \mathcal{U}_{E}^2, \ldots, \mathcal{U}_{E}^K\}$, based on their retention scores. The retention score can be computed using various metrics, such as the average number of active days per month or the lifetime value (LTV) of the user. The choice of the stratification metric and the number of levels $K$ can be determined based on domain knowledge and data analysis. For example, users can be divided into levels based on quantiles of the retention score distribution.

\subsubsection{Behavior Cloning for Each Expert Level}
For each expert level $k$, we learn a specialized behavior cloning policy $\pi_k$ that mimics the actions of experts in that level. The policy $\pi_k$ maps the user state $s_t$ to a recommendation action $a_t$, i.e., $\pi_k: \mathcal{S} \rightarrow \mathcal{A}$.

The behavior cloning objective for expert level $k$ is to minimize  the negative log-likelihood of the expert actions given the corresponding user states:
\begin{equation}
\mathcal{L}_{\text{BC}}^k = \sum_{(s, a) \in \mathcal{D}_k}-\log \pi_k(a|s),
\end{equation}
where $\mathcal{D}_k = \{(s, a)\}$ is the dataset of state-action pairs collected from the interaction trajectories of expert users in level $k$.

We parameterize the policy $\pi_k$ using a neural network architecture that consists of two main components: a state encoder $f_{\phi}$ and an action predictor $g_{\theta_k}$. The state encoder $f_{\phi}$ maps the user state $s_t$ to a compact representation $h_t = f_{\phi}(s_t)$, capturing the relevant features for decision making. The action predictor $g_{\theta_k}$ takes the state representation $h_t$ as input and generates a probability distribution over the action space $\mathcal{A}$.

For continuous action spaces, the action predictor can be modeled using a Gaussian distribution, and the behavior cloning objective can be formulated as minimizing the mean squared error (MSE) between the predicted and expert actions:
\begin{equation}
\mathcal{L}_{\text{BC}}^k  = \sum_{(s, a) \in \mathcal{D}_k} \| g_{\theta_k}(f_{\phi}(s)) - a \|^2.
\end{equation}

For discrete action spaces, the action predictor can be modeled using a categorical distribution, and the behavior cloning objective becomes the cross-entropy loss:
\begin{equation}
\mathcal{L}_{\text{BC}}^k = \sum_{(s, a) \in \mathcal{D}_k} -\log g_{\theta_k}(a|f_{\phi}(s)).
\end{equation}

The state encoder $f_{\phi}$ is shared across all expert levels, allowing for efficient learning and generalization. The action predictors $g_{\theta_k}$ are specific to each expert level, enabling specialized policies for different retention levels.

\subsubsection{Action Entropy Regularization}
\begin{figure}[t]
    \centering
    \includegraphics[width=0.48\textwidth]{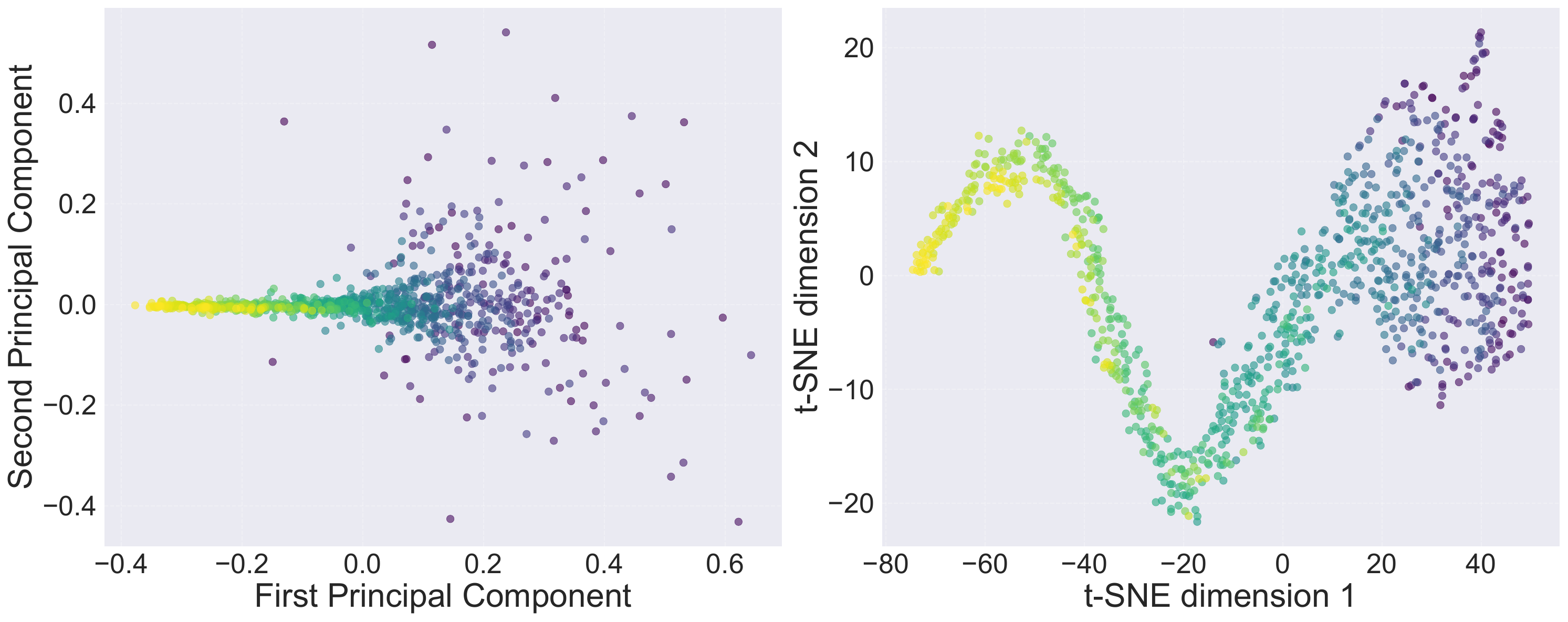}
    \caption{Visualization of the learned action distribution of a single expert policy in the principal component space (left) and t-SNE embedding space (right) using the base behavior cloning approach. The narrow concentration of actions in both projections reveals a lack of diversity in the recommended items. This limited coverage of the action space may hinder the policy's ability to capture diverse user preferences, motivating the need for entropy regularization to encourage more diverse and exploratory recommendations.}
    \label{fig:base_action}  
    \vspace{-0.2cm}
\end{figure}

Figure \ref{fig:base_action} illustrates the issue of limited diversity in the recommendation policy learned from behavior cloning. The t-SNE visualization of the action embeddings generated by the policy reveals a highly concentrated distribution. This lack of diversity can hinder the policy's ability to capture users' diverse interests and adapt to their evolving preferences. To encourage recommendation diversity and exploration, we introduce an Action Entropy Regularization (AER) term in the training objective. AER prevents the learned policies from collapsing to a narrow set of actions and promotes the discovery of novel and relevant recommendations. 

In the case of continuous action spaces, such as when the recommendations are generated in the form of item embeddings, inspired by MMCR \cite{mmcr}, we propose to use the nuclear norm of the action matrix as the entropy. Let $A_k \in \mathbb{R}^{B\times d}$ denote the action matrix from $\pi_k$, where $B$ is batch size and $d$ is the embedding dimension. Then the  regularization
term is 
$$
\mathcal{L}_{\text{AER}}^k  = -|A_k|^* = -\sum_{i=1}^{\min(B,d)} \sigma_i(A_k)
$$
where $\sigma_i(A_k)$ denotes the $i$-th singular value of $A_k$.

In the case of discrete action spaces, where the recommendations are generated as categorical items, we compute the average probability for each class and maximize the entropy of the class probabilities:
$$
\mathcal{L}_{\text{AER}}^k = \sum_{j=1}^C p_{k,j} \log p_{k,j}
$$
where $p_{k,j}$ is the average probability of class $j$ for the $k$-th expert, and $C$ is the total number of classes.

\subsubsection{Training Objectives}
The overall training objective of the SEC framework is a combination of the behavior cloning loss and the action entropy regularization term, summed over all expert levels:
\begin{equation}
\mathcal{L} = \sum_{k=1}^K \left( \mathcal{L}_{\text{BC}}^k + \lambda \mathcal{L}_{\text{AER}}^k \right),
\end{equation}
where $\lambda$ is a hyperparameter that controls the strength of the regularization.

During training, the parameters of the state encoder $f_{\phi}$ and the action predictors $g_{\theta_k}$ are optimized using stochastic gradient descent (SGD) or its variants, such as Adam.

\begin{figure}[t]
    \centering
    \includegraphics[width=0.48\textwidth]{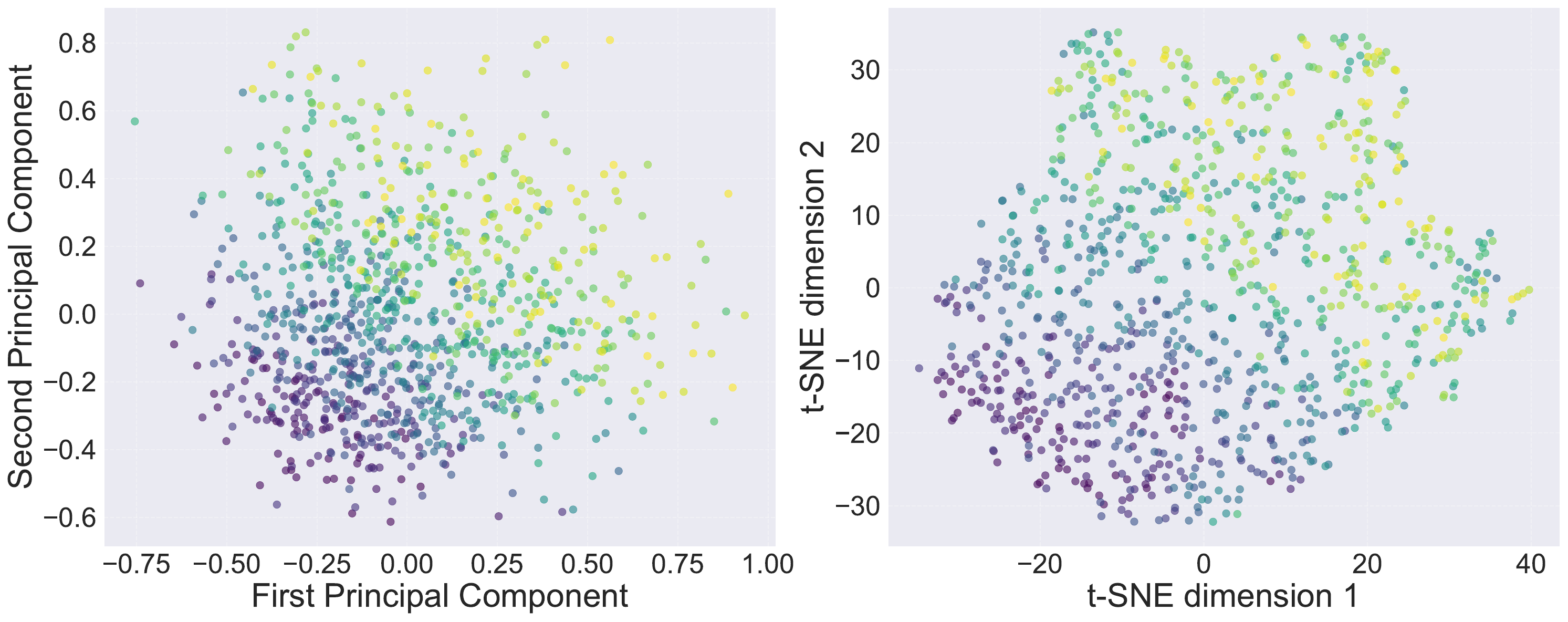}
    \caption{Visualization of action distributions of one expert in the principal component space (left) and t-SNE embedding space (right) with action entropy regularization. The regularization promotes a more diverse set of actions compared to base behavior cloning.}
    \label{fig:mmcr_action}
    \vspace{-0.2cm}
\end{figure}

\subsection{Inference Stage}
During the inference stage, the SEC framework leverages the trained expert-level policies and the adaptive selection mechanism to generate personalized recommendations for each user.
\subsubsection{User State Representation and Clustering}
To capture the representative user states for each expert level,
we apply the K-means clustering algorithm to partition the compact state space from the state encoder $f_{\phi}$ of each expert level into $C_k$ clusters. The centroid of each cluster represents a typical user state for that expert level. Let $\mu_{k,c}$ denote the centroid of the $c$-th cluster in the $k$-th expert level.
\subsubsection{Adaptive Expert Selection Mechanism}
As shown in the right part of Figure~\ref{fig:method}, The adaptive expert selection mechanism dynamically assigns users to the most appropriate expert level based on their current state and historical interactions. This allows the recommender system to adapt to the evolving preferences and retention patterns of each user.
Given a user state $s$, the adaptive selection mechanism computes the minimum distance between the user state and the cluster centroids of each expert level:
\begin{equation}
d_k = \min_c d(f_\phi(s), \mu_{k,c}), \quad k = 1, 2, \ldots, K
\end{equation}
The user is then assigned to the expert level $k^*$ that satisfies:
\begin{equation}
k^* = \text{argmin}_k \{k | d_k \leq \delta_k\}
\end{equation}
where $\delta_k$ is a predefined threshold for each expert level. If the user state does not match any expert level, the user is assigned to the expert level with the minimum distance.
To incorporate the user's historical retention information, we introduce a constraint on the selected expert level. Let $r_h$ denote the user's historical retention level, estimated based on their past interactions and retention metrics. We restrict the selected expert level $k^*$ to be no lower than the historical retention level $r_h$:
\begin{equation}
k^* = \min(k^*, r_h)
\end{equation}
This constraint ensures that the selected expert level aligns with the user's historical retention patterns and prevents the recommendation quality from degrading.

\subsubsection{Generating Retention-Oriented Recommendations}

During the inference stage, the SEC framework generates personalized recommendations optimized for user retention. Once the most suitable expert level $k^*$ is determined for a user based on their current state and historical interactions, the corresponding expert policy $\pi_{k^*}$ is applied to generate recommendations.

The expert policy $\pi_{k^*}$ takes the user state $s$ as input and outputs a set of recommended items or actions that are aligned with the retention strategies learned from the high-retention expert users at level $k^*$. By mimicking the actions of expert users who have demonstrated strong retention patterns, the policy aims to generate recommendations that are likely to keep the user engaged and promote long-term retention.

By adaptively selecting the most appropriate expert policy based on the user's state and generating retention-oriented recommendations, the SEC framework tailors its recommendations to the specific preferences and retention patterns of each user. This personalized approach aims to improve long-term user retention and satisfaction by providing recommendations that are most likely to keep the user engaged and active on the platform.

\section{Experiments}

We evaluate SEC through offline experiments and online A/B tests on two major video platforms with hundreds of millions of daily active users. During our offline testing, continuous actions are utilized, whereas discrete actions are used for the online experiments.
We aim to answer the following research questions:

\begin{itemize}[leftmargin=*]
   \item \textbf{RQ1:} How does SEC compare with state-of-the-art methods in offline evaluations? 
   \item \textbf{RQ2:} What are the effects of each component in SEC?
   \item \textbf{RQ3:} How does SEC perform in terms of user retention in online settings?
   \item \textbf{RQ4:} Can SEC be extended to optimize other long-term values?
\end{itemize}

\subsection{Offline Experiments}

\begin{table}[t]
    \centering
    \caption{Overall Performance on KuaiRand-Pure dataset.}
    \vspace{-0.3cm}
    \resizebox{\linewidth}{!}{
        \begin{tabular}{lcccc}
            \toprule
            Model & Return Time $\downarrow$ & Click Rate $\uparrow$ & Long View Rate $\uparrow$ & Like Rate $\uparrow$ \\
            \midrule
            TD3 & 2.382 & 0.800 & 0.791 & 0.852 \\
            SAC & 2.373 & 0.801 & 0.795 & \underline{0.857} \\  
            DIN & 1.947 & 0.773 & 0.764 & 0.812 \\
            CEM & 1.889 & 0.762 & 0.757 & 0.804 \\
            RLUR & 1.786 & 0.789 & 0.778 & 0.831 \\
            GFN & \underline{1.496} & \underline{0.805} & \underline{0.794} & \textbf{0.885} \\
            \hline 
            SEC(Ours) & \textbf{1.411} & \textbf{0.833} & \textbf{0.825} & 0.806\\
            \bottomrule
        \end{tabular}
    }
    \label{tab:offline_performance}
    \vspace{-0.3cm}
\end{table}

\subsubsection{Experimental Setup} 
For offline evaluations, we utilize the KuaiRand dataset, a sequential recommendation dataset with random video exposures. It includes user feedback signals such as 'click', 'view time', 'like', 'comment', 'follow', 'forward', 'hate', and 'leave'. The dataset comprises interactions from 27,285 users on 7,551 items (1,436,609 interactions, 0.70\% density).

We employ the KuaiSim retention simulator~\cite{kuaisim} to emulate long-term user behavior, which includes a leave module predicting session exits and a return module estimating platform revisit likelihood. The configurations follow the GFN model settings~\cite{gfn}.

In our experiments, users with return times of <=3 days are considered experts at different levels. We use the Policy network framework from GFN with a Transformer architecture. User embeddings are pre-trained and fixed. During inference, we set 1000 clusters per expert level with $\delta_k$ as half the mean inter-cluster distance. Optimization uses Adam with batch size 128, $\lambda=0.01$, and learning rate 0.0003.

\subsubsection{Baselines and Metrics}
We benchmark against CEM~\cite{rubinstein2004cross}, DIN~\cite{zhou2018deep}, TD3~\cite{fujimoto2018addressing}, SAC~\cite{haarnoja2018soft}, RLUR~\cite{cai2023reinforcing}, and GFN~\cite{gfn}. We evaluate using:
\begin{itemize}[leftmargin=*]
\item \textbf{Return Time}: Average days until platform return (lower is better)
\item \textbf{Click Rate}: Clicks per recommendation (higher is better)
\item \textbf{Long View Rate}: Views exceeding duration threshold (higher is better)
\item \textbf{Like Rate}: Likes per recommendation (higher is better)
\end{itemize}

\begin{table}[t]
\caption{Ablation Study Results on KuaiRand-Pure dataset.}
\vspace{-0.2cm}
\label{tab:ablation}
\centering
\resizebox{\linewidth}{!}{
\begin{tabular}{lccc}
\hline
Variant & Return Time$\downarrow$ & Click Rate $\uparrow$ & Long View Rate $\uparrow$\\
\hline
SEC (Full) & \textbf{1.411} & 0.833 & \textbf{0.825} \\
w/o Multi-level & 1.423 & 0.821 & 0.815 \\
w/o Action Entropy & 1.449 & \textbf{0.836} & 0.805 \\
\hline
\end{tabular}
}
\end{table}

\begin{figure}[t]
    \centering
    \includegraphics[width=0.45\textwidth]{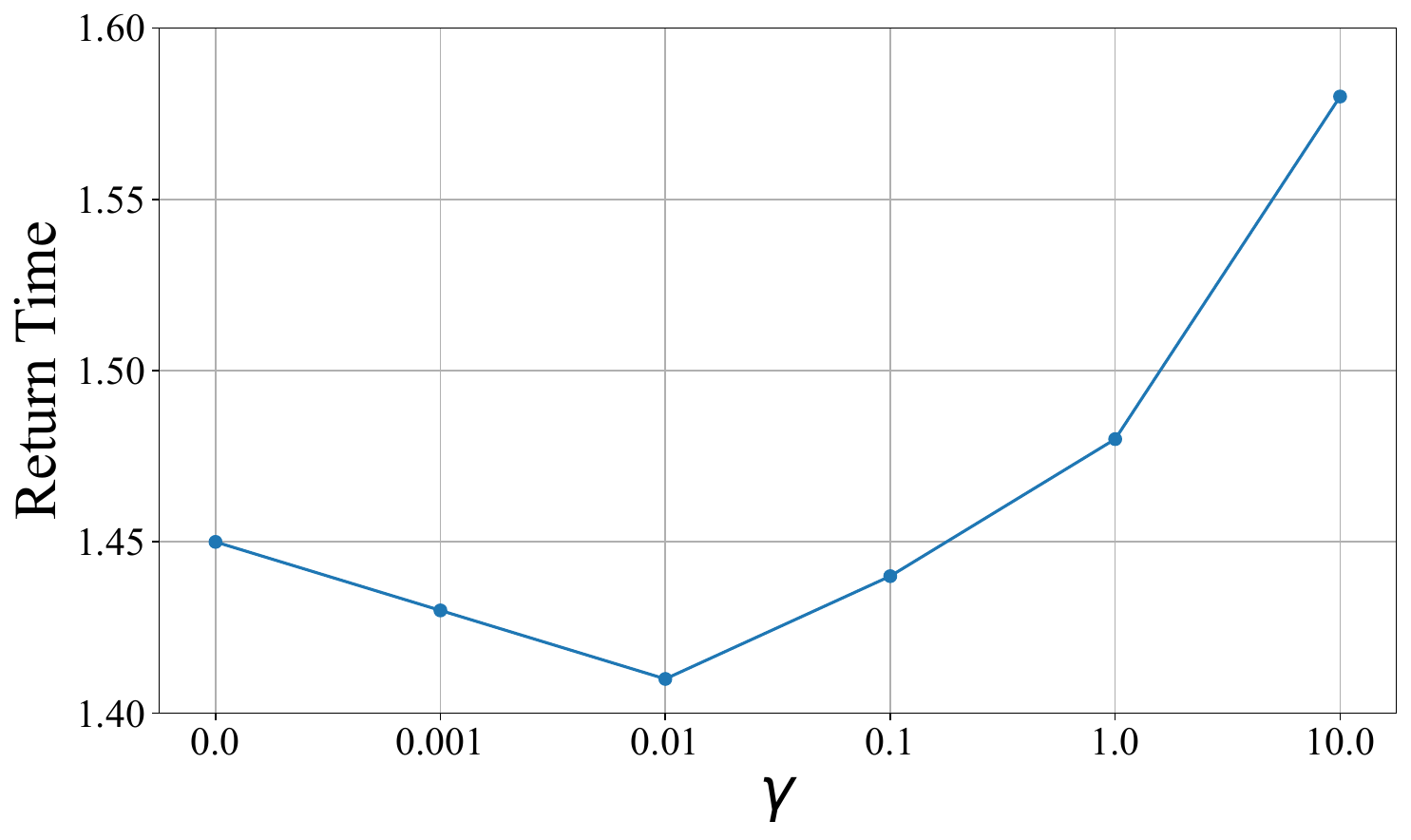}
    \vspace{-0.3cm}
    \caption{The performance of SEC with different balance loss factors $\gamma$ on the KuaiRand Dataset for minimizing return time.}
    \label{fig:balance_factor}
    \vspace{-0.3cm}
\end{figure}

\subsubsection{\textbf{RQ1 Results}}
Table~\ref{tab:offline_performance} shows SEC outperforms baselines across key metrics. For Return Time, SEC achieves a 5.7\% reduction compared to GFN, indicating improved user retention. SEC also surpasses GFN in Click Rate and Long View Rate by 3.5\% and 3.9\% respectively, demonstrating enhanced user engagement.

However, SEC's Like Rate is 8.9\% lower than GFN's, reflecting a trade-off between immediate satisfaction and long-term retention. This aligns with SEC's primary objective of optimizing user retention rather than maximizing immediate satisfaction.

\begin{table*}[t]
\centering
\caption{Online A/B test results for User Retention. Bold values indicate statistically significant results (p-value < 5\%).  The total gains translate to over 200,000 additional daily active users (DAU) for each platform.}
\vspace{-0.3cm}
\label{tab:retention_results}
\begin{tabular}{llccc}
\toprule
\textbf{Metric} & \textbf{Model} & \textbf{Stage} & \textbf{Kuaishou} & \textbf{Kuaishou Lite} \\
\midrule
\multirow{4}{*}{Active Days} & \multirow{2}{*}{Behavior Cloning + Entropy Loss} & 1st Ranking & \textbf{+0.034\%} & \textbf{+0.037\%} \\
 & & 2nd Ranking & \textbf{+0.036\%} & \textbf{+0.042\%} \\
\cmidrule{2-5}
 & Additional gain from Multi-level Expert & Both Stages & \textbf{+0.028\%} & \textbf{+0.043\%} \\
\cmidrule{2-5}
 & \textbf{Total Improvement} & & \textbf{+0.098\%} & \textbf{+0.122\%} \\
\bottomrule
\end{tabular}
\vspace{-0.1cm}
\end{table*}

\subsubsection{\textbf{RQ2 Results}}
Our ablation study (Table~\ref{tab:ablation}) shows removing multi-level behavior cloning increases Return Time by 0.9\% and decreases Click Rate and Long View Rate by over 1.0\%, demonstrating the importance of learning from experts at different retention levels.

Removing action entropy regularization increases Return Time by 1.8\% and Click Rate by 1.7\%, but decreases Long View Rate by 2.4\%. This suggests that while entropy regularization may reduce immediate engagement, it enhances sustained engagement.

Figure~\ref{fig:balance_factor} shows SEC's performance with different balance factors. The return time is optimized at moderate values (around 0.01), while very high or low values lead to performance degradation. This highlights the importance of balancing behavior cloning and action entropy regularization.

\subsection{Online Experiments}
\begin{figure}[t]
    \centering
    \includegraphics[width=0.4\textwidth]{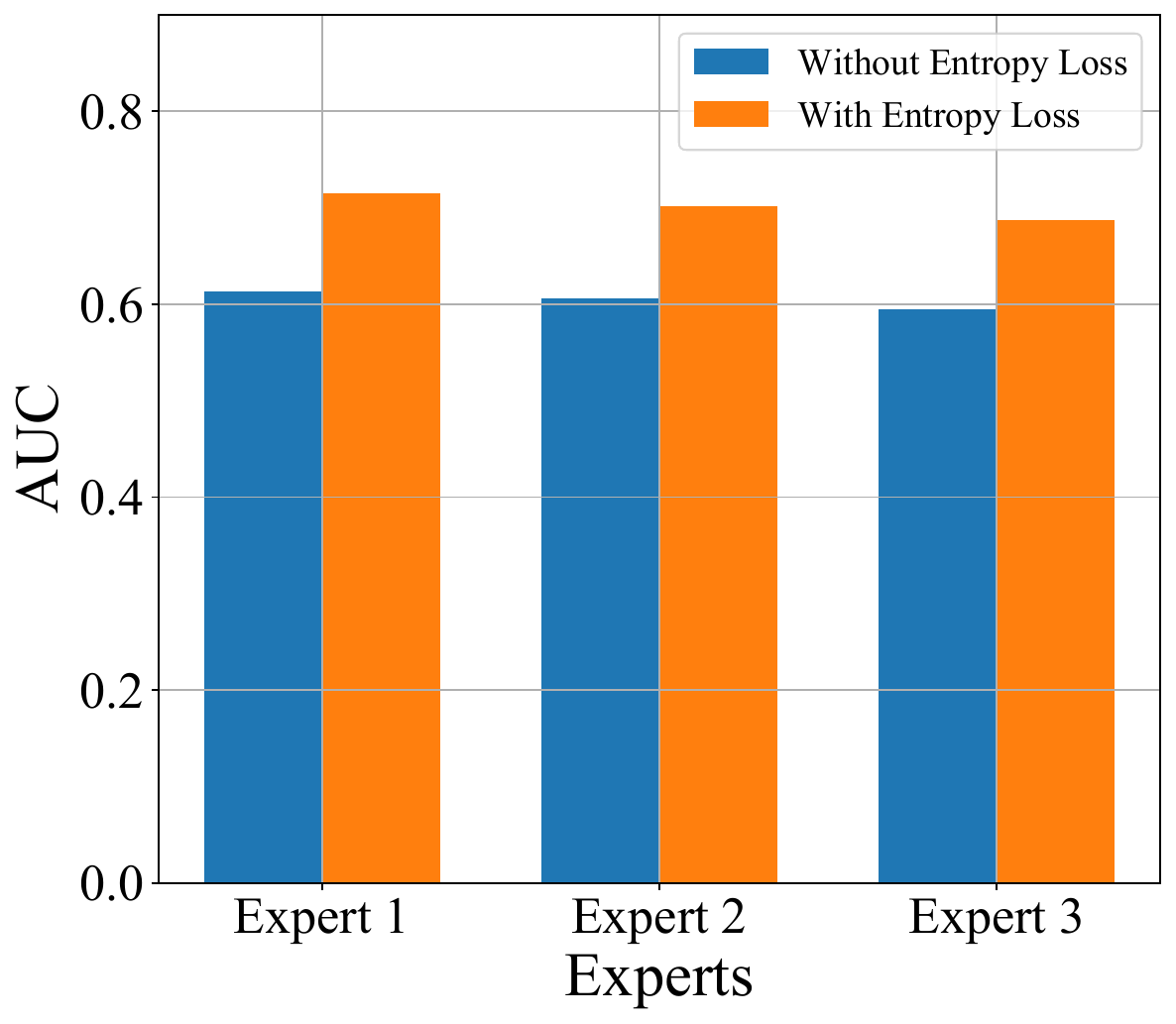}
    \vspace{-0.3cm}
    \caption{AUC scores during online experiments with and without action entropy regularization across different expert activity levels.}
    \label{fig:online_auc}
    \vspace{-0.3cm}
\end{figure}
\subsubsection{Experimental Setup}We deployed SEC on Kuaishou and Kuaishou Lite platforms, each with over 200 million daily active users. Unlike offline experiments where our strategy controlled item ranking completely, in online environments, we integrate with existing ranking systems. 

Due to the vast scale of video IDs, we used video clustering based on multimodal features to generate recommendation actions. 
We combined cluster action prediction probability with other fine-grained ranking scores in the two ranking stages to rank videos.
We employed MLP networks for both State Encoder and Action Predictor, with users having high activity days classified into 3 expert levels. In our A/B tests, we allocated 10\% traffic to both baseline and SEC-optimized versions for over two weeks.

\subsubsection{\textbf{RQ3 Results}} 
Table~\ref{tab:retention_results} shows SEC significantly improved Active Days (users active within a 7-day window) across both platforms. First, we implemented Behavior Cloning + Entropy Loss, achieving +0.034\% and +0.037\% gains in 1st Ranking stage, and +0.036\% and +0.042\% in 2nd Ranking stage for Kuaishou and Kuaishou Lite respectively.

Adding Multi-level Expert stratification yielded additional gains of +0.028\% and +0.043\%. The cumulative improvements (+0.098\% for Kuaishou, +0.122\% for Kuaishou Lite) translate into over 200,000 additional daily active users on each platform, underscoring SEC’s real-world impact at scale.

Additionally, we examine the impact of incorporating action entropy regularization into our training process. The results indicate that this addition improves AUC scores during online training, with varying levels of improvement observed across different activity levels of experts, as shown in Figure~\ref{fig:online_auc}. Notably, our AUC calculations are based on a dataset containing billions of samples. This suggests that our action entropy regularization and behavior cloning loss are isotropic, effectively enhancing overall model performance.

\subsubsection{\textbf{RQ4 Results}} 
We also evaluated SEC's impact on user interest expansion (Table~\ref{tab:interest_expansion}). Valid interest clusters (video clusters users frequently interact with) increased by 1.31\% on Kuaishou and 1.14\% on Kuaishou Lite, demonstrating SEC's ability to broaden user interests.

These results highlight SEC's practical value in industrial-scale recommendation systems, showing improvements in both user retention and interest diversity through learning from experts at different retention levels, adaptive selection, and encouraging recommendation diversity.

\begin{table}[t]
\centering
\caption{Online A/B test results for User Interest Expansion. Bold values indicate statistically significant results ($p$-value $< 5\%$).}
\vspace{-0.3cm}
\label{tab:interest_expansion}
\begin{tabular}{lccc}
\toprule
\textbf{Metric} & \textbf{Stage} & \textbf{Kuaishou} & \textbf{Kuaishou Lite} \\
\midrule
Valid Clusters & 1st Ranking & \textbf{+1.31\%} & \textbf{+1.14\%} \\
\bottomrule
\end{tabular}
\vspace{-0.3cm}
\end{table}

\section{Limitations}
While the proposed SEC framework demonstrates strong performance, its applicability is constrained by several limitations. First, its effectiveness is fundamentally contingent on a sufficiently large and diverse set of high-retention "expert" users, which may not be available on nascent platforms or in niche domains. Furthermore, the adaptive expert selection mechanism relies on high-quality user state representations; these can be sparse or unreliable for new users, posing a significant challenge in cold-start scenarios. Lastly, although validated on short-video platforms, the generalizability of SEC to other domains such as e-commerce or news recommendation warrants further investigation, as the definitions of expert behavior and retention metrics can vary significantly across contexts.

\section{Conclusion}

In this work, we introduced Stratified Expert Cloning (SEC), a novel imitation learning framework designed to optimize long-term user retention in large-scale recommender systems. SEC integrates three key innovations: multi-level expert stratification to capture diverse retention behaviors, adaptive expert selection to match users with appropriate policies, and action entropy regularization to enhance recommendation diversity. Our comprehensive evaluation, featuring extensive offline experiments and large-scale A/B tests, confirms SEC's superior performance over state-of-the-art methods, with ablation studies validating the contribution of each component. By circumventing the exploration risks and delayed reward challenges inherent in reinforcement learning, SEC offers a practical and effective approach to building recommender systems that deliver long-lasting user value at scale.

\bibliographystyle{ACM-Reference-Format}
\balance
\bibliography{sample-base}

\appendix 
\section{GenAI Usage Disclosure}
During the preparation of this work, GenAI tools were used solely to improve the spelling and grammar of the author-written text. The authors are fully accountable for the content.


\end{document}